\documentclass{article}
\usepackage{graphicx}
\usepackage{authblk}
\usepackage{verbatim}
\usepackage{array, tabularx}
\usepackage{subfigure}
\usepackage{caption}

\usepackage[a4paper, total={6.6in, 8in}]{geometry}
\usepackage{xr}

\usepackage[utf8]{inputenc}
\usepackage[T1]{fontenc}
\usepackage{amsthm}
\usepackage{amsmath}
\usepackage{amssymb}
\usepackage{siunitx}

\usepackage{url}
\usepackage{booktabs} 
\usepackage{xspace}
\usepackage{float}
\usepackage{multirow}
\usepackage{alltt}
\usepackage{color}
\usepackage{pifont}
\usepackage{makecell}
\usepackage{multicol}
\usepackage{cleveref}
\usepackage[super]{nth}
\usepackage{xcolor}
\usepackage{colortbl}

\title{Collective Intelligence Outperforms Individual Talent: A Case Study in League of Legends}

\author[1]{Angelo Josey Caldeira}
\author[1]{Sajan Maharjan}
\author[1]{Srijoni Majumdar}
\author[1]{Evangelos Pournaras}

\affil[1]{School of Computer Science, University of Leeds, Leeds, UK, \ \ \ \ \ \ \ \ \ \ \ \ \ \ \ \ \ \ \ \ \   \ \ \ \ \ \ \ \ \ \ \ \ \ \ \ \ \ \ \ \ \ \ \ \ \   \ \ \ \ E-mails: \{sc20ajc, scsmah, s.majumdar, e.pournaras\}@leeds.ac.uk}

\begin{document}
\maketitle
	
\begin{abstract}
\footnotetext[1]{Corresponding author: Sajan Maharjan, School of Computing, University of Leeds, Leeds, UK, E-mail: scsmah@leeds.ac.uk}

Gaming environments are popular testbeds for studying human interactions and behaviors in complex artificial intelligence systems. Particularly, in multiplayer online battle arena (MOBA) games, individuals collaborate in virtual environments of high realism that involves real-time strategic decision-making and trade-offs on resource management, information collection and sharing, team synergy and collective dynamics. This paper explores whether collective intelligence, emerging from cooperative behaviours exhibited by a group of individuals, who are not necessarily skillful but effectively engage in collaborative problem-solving tasks, exceeds individual intelligence observed within skillful individuals. This is shown via a case study in \emph{League of Legends}, using machine learning algorithms and statistical methods applied to large-scale data collected for the same purpose. By modeling and visualizing systematically game-specific metrics but also new game-agnostic topological and graph spectra measures of cooperative interactions, we demonstrate compelling insights about the superior performance of  collective intelligence.

\end{abstract}

\section{Introduction} \label{sec:intro}
\maketitle
The recent advancements in computing technology have evolved the way individuals socialize and interact; online gaming platforms being one of the most prominent mediums for social interaction and cooperation. Zhong reports that in China there are 147 million people playing online games, thus representing new forms of community, social interaction, and collaboration~\cite{zhong2011effects}. These online gaming environments are popular testbeds for studying human interactions and behaviors in complex artificial environments. The simulation of diverse scenarios ranging from strategic decision making to adaptive behaviors within online gaming environments allows the testing of complex real-world mechanisms under controlled low-cost conditions. Furthermore, the virtual worlds of such online games are populated with autonomous artificial agents and other human players, thus offering opportunities for interesting interactions that guide players and enhance social dynamics~\cite{laird_human_level_2001}, resulting in a \emph{collective intelligence} derived from the synergistic cooperation of humans and computing agents. Collective intelligence is the general ability of a particular group to perform well across a wide range of different tasks~\cite{woolley2015collective}. Collective intelligence also involves the study of how groups of one person and one computer interact (human-computer interaction) as well as how larger groups of people interact with computing technologies to cooperate and make intelligent decisions, for instance, in the development of collaborative software such as Wikipedia~\cite{malone2022handbook}. 

\begin{figure}[!htbp]
    \centering
    \includegraphics[scale=0.27]{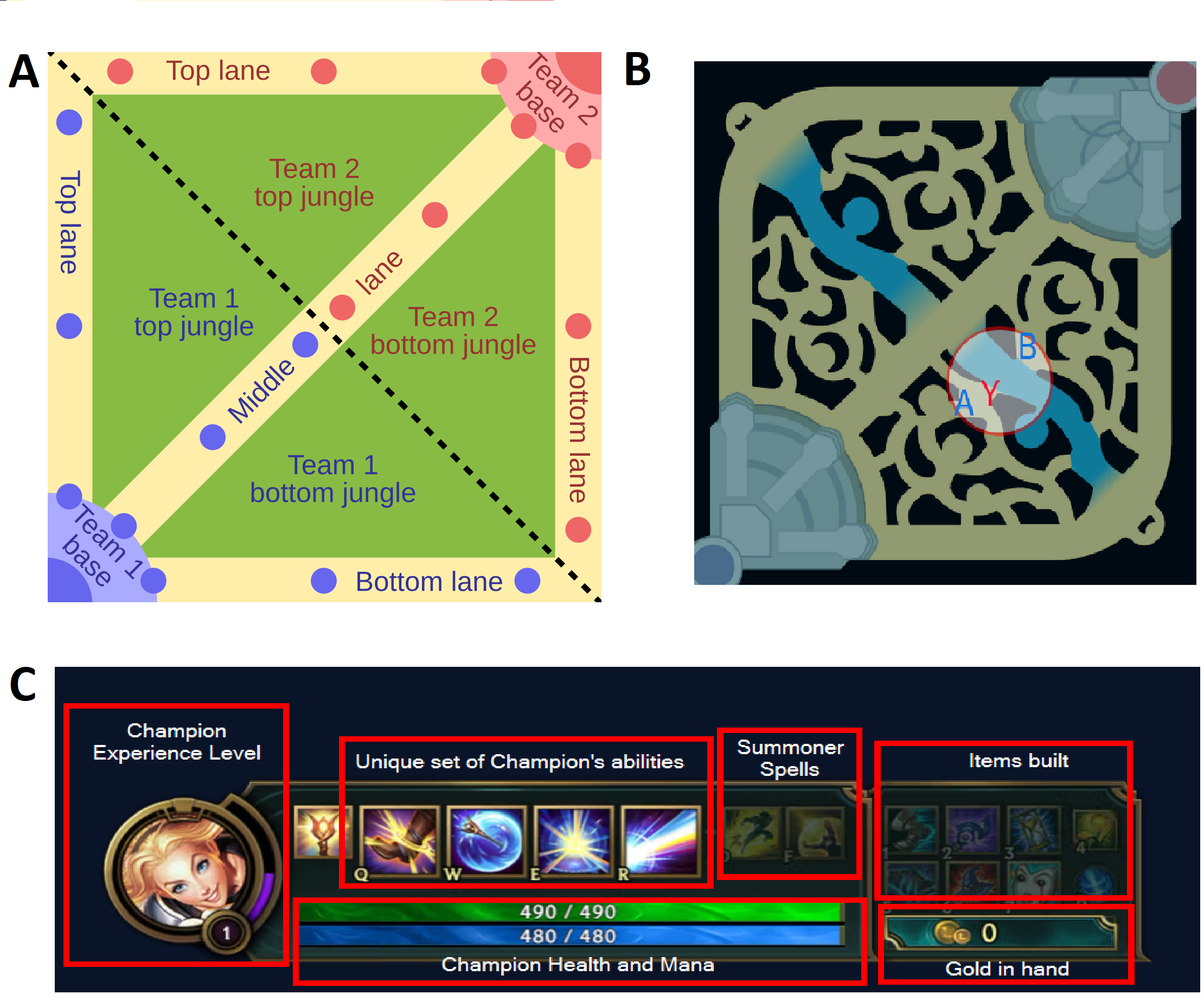}
    \caption{(A) The three lanes, 2 jungles and their subdivisions. Small red and blue circles denote tower structures for the respective teams. The darker, inner-circle segment in the top-right and bottom-left corners denote each teams’ Nexus; (B) An instance of gameplay on Summoner's Rift where an opponent kill is scored due to positional map-pressure. Blue team player A attempts to kill red team player Y, who tries to flee. Blue player B blocks Y's escape, forcing engagement and leading to A's kill. Although B does not actively assist, B’s positioning creates map pressure resulting in Y’s death. B is awarded a map-pressure assist if it is positioned within a specified radius during the kill.
(C) Details of an in-game champion;}
    \label{fig:archi_s}
\end{figure}

 MOBA games are a subgenre of real-time strategy games in which two teams, each consisting of 5 individual players, compete against each other with each player controlling a single character (called a \emph{champion}), while strategy revolves around individual character development and cooperative team play in combat~\cite{mora2018moba}. Riot Games' \emph{League of Legends} is the most popular MOBA game played worldwide with over 150 million monthly active players~\cite{gandhi2024beliefs}. \emph{League of Legends} features 171 uniquely playable champions with their own set of skills, that relies on an individual player's micro-management ability, paired with a strong understanding of champion \emph{mechanics} along with the collective effort of a team to think strategically and showcase team-work abilities. The game is structured around a map called \emph{Summoner's Rift}, and the primary objective of the game is to destroy the opponent team's base, a tower structure called the \emph{Nexus}, while the game progresses with champion development via earning gold, gaining experience, crafting items and scoring kills (refer Figure~\ref{fig:archi_s}).
 \footnote{Figure 1A is sourced from Marcal Mora Cantallops~\cite{mora2018exploring}, while Figure 1B and Figure 1C are re-worked from in-game interfaces}

The collaborative dynamics among randomly assigned players in a team, each controlling a unique champion and working toward a shared win objective, exemplify collective intelligence in MOBA games. In-game chat messages, signals, and assists used to secure objectives offer measurable indicators of collective intelligence. Leavitt et al. show that nonverbal communication like in-game pings improves team strategy and performance~\cite{leavitt2016ping}. Meanwhile, individual players may excel in champion mechanics, reflecting high individual intelligence. Extending existing work~\cite{glombik2025visually}, this  paper uses in-game data to model and visualize metrics of both individual and collective intelligence, testing the hypothesis \textbf{H1: collective intelligence outperforms individual intelligence of the players in winning games}. These players in such competitive games are categorized into a two-tiered ranking system of \emph{tier} and \emph{division}; based on recent performance, serving as proxies for individual skill levels that influence outcomes. This motivates \textbf{H2: Highly-skilled players exhibit higher levels of collective intelligence compared to less-skilled players}. Although player perceptions of in-game incentives affect skill development~\cite{richter2015studying}, such mechanisms are beyond the scope of this study. The main contributions of this work in evaluating individual and collective intelligence in \emph{League of Legends} are: (i) the first use of effective graph resistance from graph spectra as a proxy for collective intelligence, previously used for network robustness~\cite{wang2014improving}; (ii) new visual communication of collective intelligence by combining graph and complex network approaches with chord diagrams; and (iii) new insights into how collective intelligence appears in \emph{League of Legends} and how it compares with individual player intelligence.

\section{Related Work}\label{sec:related-work}
Collective intelligence can be observed in human-computer interaction, collective-decision making and collaborative software development; in the decentralized collective learning of self-managed economies~\cite{pournaras2018}, swarm intelligence in the large-scale coordination of drones~\cite{qin2023coordination}, etc. Woolley et al. first provided the evidence for a collective intelligence factor, \emph{c} in human groups, suggesting that \emph{c} is not correlated with the average or maximum individual intelligence of the group members, but rather it emerges from the way group members interact~\cite{woolley2010evidence}. Krause et al. conclude that the potential benefits of human swarm intelligence depend on the type of problem, while advocating for diversity in groups, and that high-performing individuals can be outcompeted by a similar size group of  low-performing individuals~\cite{krause2011swarm}. Likewise, Santos et al. argue that social diversity catalyzes cooperative behaviour~\cite{santos2012role}. The sports and gaming industries have integrated collective intelligence paradigms to solve problems and generate insights beyond the individual capabilities of a single player.

Duch et al. quantified the contribution of individual players towards the overall team performance during the 2008 European Cup football tournament via network graphs that modeled the flow of passes among players in a team~\cite{duch2010quantifying}. Grund confirmed that centrality of interaction between players leads to a decreased team performance by analyzing soccer passes of teams in the English Premier League~\cite{grund2012network}. Szabo et al. investigated team behavioral dynamics in \emph{escape room} settings and reported that more effective teams tend to solve problems via coordinated, balanced communication patterns, maintaining a dynamic alternation between different types of tasks~\cite{szabo2024coordination}. In the realm of digital gaming, \emph{42 Entertainment} launched \emph{I Love Bees}, an alternate reality game used for the marketing of \emph{Halo 2} in which over 600,000 players collaborated in solving puzzles related to the game prior to its release~\cite{mcgonigal2008love}. Gonzalez-Pardo et al. conducted an empirical study describing how collective intelligence algorithms like ant-colony optimization and genetic algorithms can be used for problem-solving within video games~\cite{gonzalez2015empirical}. Kokkinakis et al. explore the relationship between fluid intelligence and video game performance in MOBAs under controlled laboratory conditions, and found that fluid intelligence correlates significantly with in-game rank~\cite{kokkinakis2017exploring}.

Specifically within \emph{League of Legends}, Kim et al. have shown that teams exhibit group collective intelligence and that such measures for collective intelligence correlate with their in-game performance~\cite{kim2015work}. In addition, Kim et al. have shown that collective intelligence can predict the competitive performance of teams controlling for the amount of time that different individuals play together as a team, and that collective intelligence is positively correlated with the presence of a female team member and the average social perceptiveness~\cite{kim2017makes}. Sapienza et al. study the evolution of individual performance within ad-hoc teams and claim that player performances in successive matches deteriorates over the course of a gaming session~\cite{sapienza2018individual}. Kou and Gui report the existence of rich social interactions within temporary teams via semi-structured interviews with experienced players~\cite{kou2014playing}. Mora-Cantallops and Sicilia have identified 4 different clusters of players associated with team cohesiveness, based on a systematic classification of player-centric networks measured by a set of graph metrics~\cite{mora2018player}. 
 
Subsequent research by Mora-Cantallops and Sicilia investigate team efficiency in professional player settings through directed graph networks that indicate the flow of assists between players and argue that team efficiency is positively affected by interactions between players while centralization of resources is detrimental~\cite{mora2019team}. Likewise, Kho et al. employ neural networks to reverse engineer games played in three professional league settings and have established an induced logic across different leagues that determine the winning outcome of a game depending on different objectives scored within the game~\cite{kho2020logic}. Maymin generated high-frequency data on millions of ranked games and calibrated a live, in-game win probability model based on the conditions of the game at each moment~\cite{maymin2021smart}. Also, Ani et al. use ensemble methods to predict the outcome of a match, from a dataset of 1500 games using 97 different \emph{pre-match} and \emph{within-game} features.

The novel contribution of this research emerges from an inter-disciplinary approach that explores collective intelligence among random groups of players in the virtual world of \emph{League of Legends}, by using different machine learning methods to predict the winning outcome of games. That is, the novelty of this work does not lie in the sophisticated design of machine learning models to better predict the winning outcome of a game, but rather quantifies and contrasts the importances of individual and collective intelligence behaviours among players via an empirical analysis of the collected data. The data-driven analysis presented in this work reveals an interesting insight that explains the nature of individual and collective team behaviors i.e. \emph{acquiring} vs \emph{sharing} and \emph{cooperative} vs \emph{non-cooperative}; and necessitates the importance of altruistic behaviors. 

\section{Methods}\label{sec:methods}
\begin{figure}[!htp]
    \centering
    \includegraphics[width=\columnwidth]{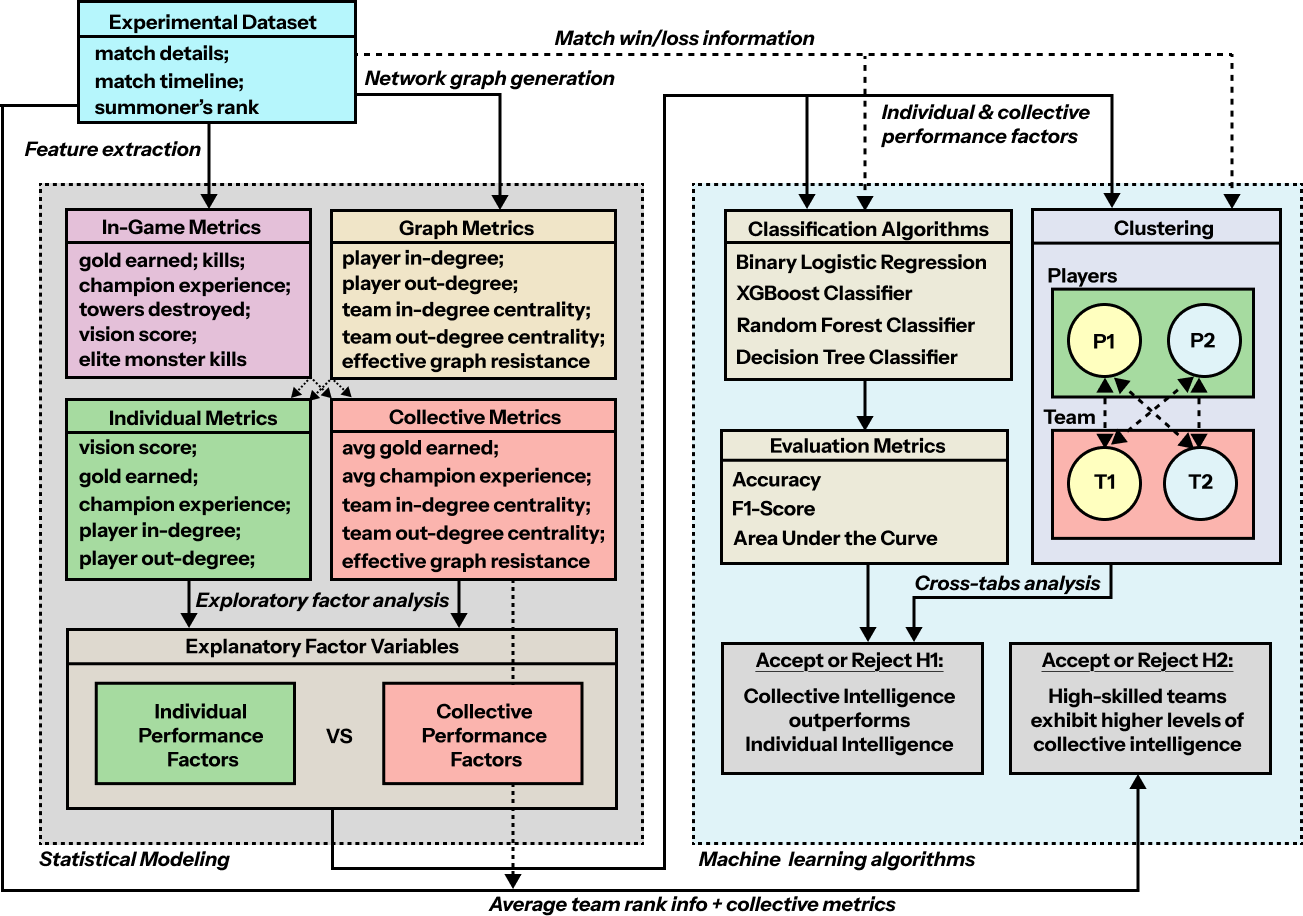}
    \caption{ Overall architectural framework for the evaluation of collective intelligence hypotheses within \emph{League of Legends} gameplay environments.\textnormal{ The architecture comprises two main components: (i) computation of individual and team metrics, and (ii) analysis of winning outcomes using exploratory data analysis with classification and clustering methods on such individual and collective team-level data.}}
    \label{fig:lol_evaluation_architecture}
\end{figure}

To evaluate the aforementioned hypotheses within the \emph{League of Legends} gameplay environment, different statistical methods and machine learning algorithms are implemented. Figure~\ref{fig:lol_evaluation_architecture} depicts the overall framework of the methods used in this study. Individual in-game performance metrics are extracted from the experimental dataset while team-based collective intelligence metrics are computed from their corresponding directed graph networks that outline the flow of resources between players in a team. Exploratory factor analysis is applied to both individual and team-level data, revealing latent performance factors. Such latent factors are input as features to predict the winning outcome of games using different classification algorithms in order to understand the nature of such intelligent behaviours. Comparison of the classification results based on the individual and collective performance factors determine whether or not to accept the hypothesis \textbf{H1}. Furthermore, k-means clustering is used to profile players and teams into different clusters. Crosstabs analysis of the winning rate of different teams having different player clusters reinforces whether or not to accept the hypothesis \textbf{H1}. Finally, statistical hypothesis testing on \textbf{H2} is employed to check for the significance in the difference of collective intelligence levels of teams (and players) across different ranks in the case of team wins.

\subsection{Experimental Dataset} 
\emph{League of Legends} game data played on official servers is publicly accessible via a set of RESTful APIs provided by Riot Games~\cite{riotgamesAPI}. In order to fetch game data for competitive matches played by a given player, a \emph{summonerName} which uniquely identifies the player, is required. A preliminary list consisting of 10 random \emph{summonerName}s corresponding to the 10 different ranking tiers, is used as initial seed data. For each of these seed \emph{summonerName}s, the API offered by Riot Games is used to retrieve -- (a) a list of $N=100$ recent matches played by the player, and for each of these matches, (b) detailed match data that include game outcome and individual match performance of the players, (c) a list of game events that occurred at periodic time intervals within the match and, (d) public rankings of the associated players. The API fetches not just the match performance of the seeded \emph{summonerName} player, but also retrieves details for the other 9 players relevant in a given match. An additional set of data i.e., the most recent $N=10$ games for all the relevant players identified from the initial fetch, is collected, thus removing any initial selection bias. Using the process described above, a total of 103K matches are fetched. However, due to different modes of gameplay offered in the game, the collected dataset does not exclusively consist of ranked competitive matches only. Furthermore, two different players in this dataset may have played in the same game, or there can be games where a player was disconnected (due to Internet connection issues, or abandoning  the game) from the server resulting in less than 10 players, or the game was terminated early via a \emph{remake} or a \emph{surrender}. To remedy such issues, the dataset was filtered to remove duplicate entries, retrieving only those games in which all 10 players were present throughout the game in a competitive ranked match mode that was not terminated early (i.e., game duration is not less than 23 minutes preventing any early surrender and ensuring both teams are competing to win towards the latter stages of the game regardless of the early dominance by a team). Thus the experimental dataset consisted a total of 31,055 ranked matches with 164,571 unique players. For each game, in-game performance metrics of all players were extracted and directed graph networks were generated to formulate team-based graph metrics.

\subsection{In-game Performance Metrics}
The dataset collected from the API endpoints, particularly the \emph{match detail} endpoint, consists of fields that reflect the respective in-game performance of all players in a match in terms of metrics such as -- \emph{experience gained, gold earned, kills score, assists given, champion deaths, vision score, elite monsters kills, towers destroyed, minions killed}, etc. along with the overall match outcome. Feature scaling was performed to resolve the discrepancies in the range of values for the in-game metrics. Additionally, per-minute metrics for \emph{gold earned, assists given, vision score, experience gained} and \emph{minions killed} were also computed to scale and normalize the range of features in the dataset. Additionally, champions played by individual players feature their own set of unique \emph{skillshots, spells} and \emph{items} crafted during the gameplay which can impact the outcome of a game. However, incorporating such features only adds complexity without providing high-level insights on the nature of intelligent behaviors. Thus champion \emph{skillshots, spells} and \emph{item builds} are not accounted as in-game performance metrics for simplicity.

\subsection{Graph Metrics}
Graph metrics characterize the flow and centrality of resources between individuals in a team. Such graph metrics signify how teams and individuals interact, strategize and cooperate throughout the game to secure different in-game objectives and resources. To explore the interaction among players, we utilize the \emph{match timeline} data to fetch game events during which players in a team assisted one another to secure objectives and resources such as opponent kills, enemy tower destruction, elite monster kills as well as unrecorded assists for \emph{positional map-pressure} resulting in a kill. \emph{Positional map-pressure} refers to the intelligent positioning of a player on the map to prevent an opponent player from escaping, forcing its kill without actively participating in it. 

Based on the flow of these assistance between players in a team, directed graph networks with 5 nodes (signifying the 5 players) with directed, weighted edges corresponding to the flow of assistance are created for both teams in each game. For any such graph network, let $N=5$ be the total number of nodes, $A$ be the total observed assists between players in the team, and $w_{ij}$ be the weight of directed edge (i.e. the number of assists) from node $i$ to node $j$. Then, the following graph metrics are computed --   \\ \textbf{1. Player In-Degree:} This metrics signifies the weighted sum of assists received by a player from all other players in the team i.e. the in-degree of player $i$ is: $C_{ID}(i) = \sum_{j=1}^{N} w_{ji}$.  \\ \textbf{2. Player Out-Degree:} This metric signifies the weighted sum of assists given by a player to all other players in the team i.e. the out-degree of player $i$ is: $C_{OD}(i) = \sum_{j=1}^{N} w_{ij}$. \\ \textbf{3. Team In-Degree Centrality:} This metric signifies the extent to which assists are centralized towards any player in a given team i.e.: $C_I = \frac{\sum_{i=1}^{N} (C_{ID}^{max} - C_{ID}(i))}{(N -1)A}$. \\ \textbf{ 4. Team Out-Degree Centrality:} This metric signifies the extent to which assists are given out by any player in a given team i.e.: $C_O = \frac{\sum_{i=1}^{N} (C_{OD}^{max} - C_{OD}(i))}{(N -1)A}$. \\ \textbf{5. Effective Graph Resistance (EGR):} This metric measures the robustness of a graph network~\cite{wang2014improving}. A higher value of this metric signifies the lack of cooperation among nodes in the graph network. This metric is computed by extracting eigenvalues ($\mu$) from the corresponding Laplacian matrix of the graph network for each team i.e.: $EGR = N \sum_{i=1}^{N-1} \frac{1}{\mu_{i}}$. 

\begin{figure}[!htbp]
    \centering
    \includegraphics[width=0.5\textwidth]{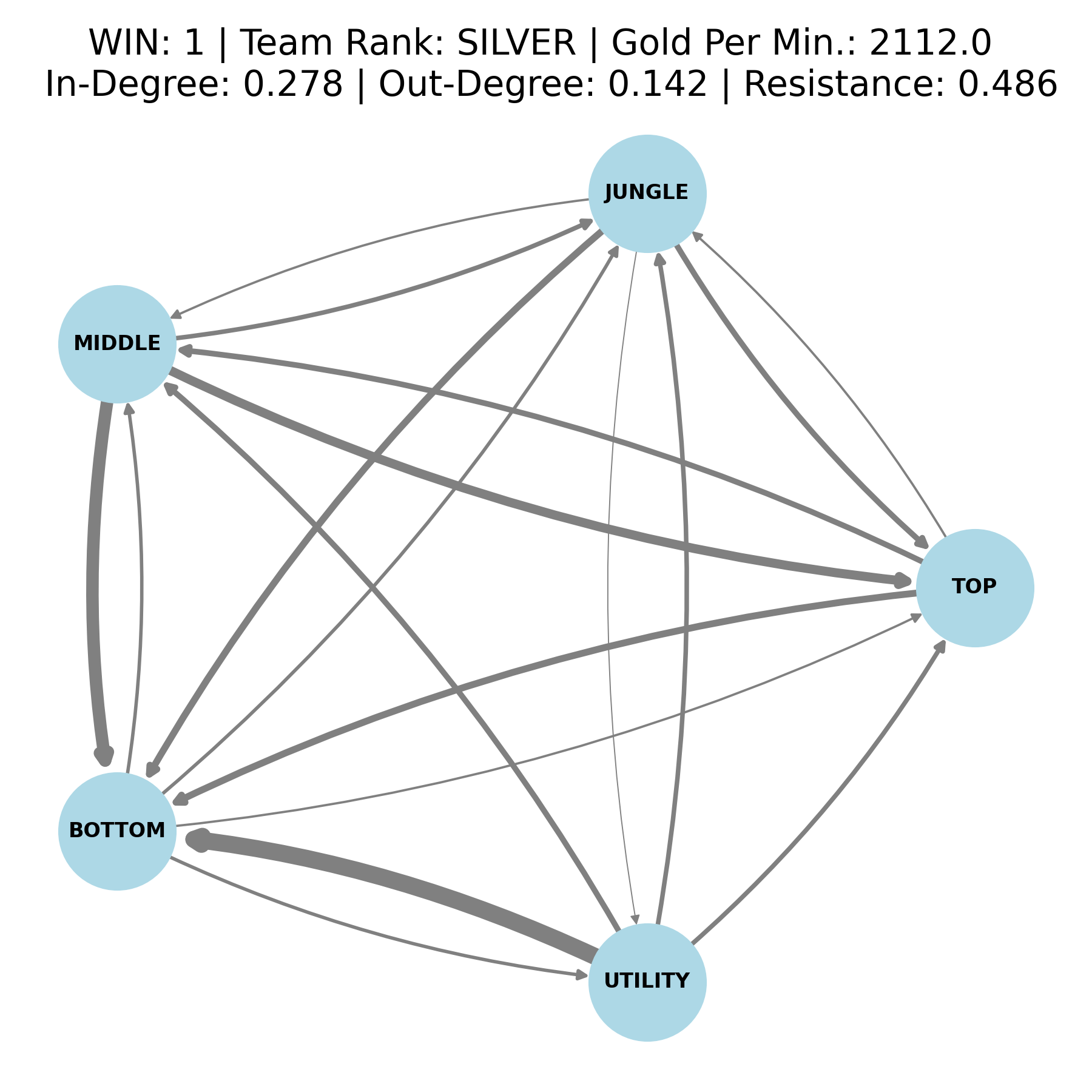}
    \caption{Directed graph network for a sample team characterizing the flow of interactions between players. Each graph has 5 nodes corresponding to player roles and weighted, directed edges signify the frequency and flow of all forms of assistance from one player to another. Average values of team performance and computed graph metrics is shown.}
    \label{fig:network_graphs}
\end{figure}

The graph metrics -- \emph{player in-degree, player out-degree, team in-degree centrality, team out-degree centrality} have been used in previous work~\cite{mora2019team}, but our approach models all forms of assistance i.e. assistance on \emph{kills} including \emph{towers taken, elite-monsters and positional map-pressure}

\noindent \\ A directed graph network signifying the flow of interactions among players in a given team is illustrated in Figure~\ref{fig:network_graphs}.

\subsection{Individual Metrics vs Collective Metrics}
An individual's performance in a game is accounted by the in-game performance metrics of the individual player such as \emph{gold earned, champion experience, kills}, etc. Correlation analysis and variance inflation factor tests revealed multi-collinearity between in-game metrics which were thus removed from the feature set. Additionally, the graph metrics i.e. \emph{player in-degree} and \emph{player out-degree} which signify the overall receipt and provision of all assists for a single individual player is also accounted towards individual metrics. The collective metrics are composed of average values of in-game performance metrics of the players in a team along with the graph metrics that signify the overall flow of assistance within the team. Table~\ref{tab:individual_and_collective_metrics} outlines the list of individual metrics and collective metrics used in the study.

\begin{table}[htbp!]
\centering
\caption{Performance metrics of individuals and teams}
\label{tab:individual_and_collective_metrics}
\begin{tabular}{cc}
\toprule
Individual Metrics & Collective Metrics \\
\midrule
Gold Earned Per Min. & Avg. Gold Earned Per Min. \\
Experience Per Min. & Avg. Experience Per Min. \\
Vision Score Per Min. & Avg. Vision Score Per Min. \\
Player In-degree & Team In-degree Centrality \\
Player Out-degree & Team Out-degree Centrality \\
& Effective Graph Resistance \\
\bottomrule
\end{tabular}
\end{table}

\subsection{Exploratory Factor Analysis} 
Prior work by Mora-Cantallops and Sicilia have confirmed that gold earned is one of the most important factors determining the game outcome~\cite{mora2019team}. Gold is earned by all players at a steady rate throughout the game in addition to champions and minion kills, tower takedowns and other objectives. However, only using gold earned as a feature doesn't identify the nature of intelligent behaviors exhibited by individuals and teams. Therefore, exploratory factor analysis is performed to uncover latent factor variables that explain the nature of intelligent interactions between individuals and teams. The optimal number of explanatory factors for individuals and teams are identified via Scree analysis. Factor loading values of the latent factors infer the nature of interactions in both individuals and teams i.e. \emph{acquiring} vs \emph{sharing} factor and \emph{cooperative} vs \emph{non-cooperative} factor.

\subsection{Machine Learning Algorithms} 
\textbf{Classification:} A game's winning outcome is often driven by either exceptional individual performance or collective team effort. Explanatory factors reflecting intelligent behaviors in individuals and teams are used as features to predict the game's outcome through classification algorithms—binary logistic regression, decision tree, random forest, and XGBoost. Results are evaluated using accuracy, f1-score, and area under the curve. To assess \textbf{H1}, evaluation scores from collective performance factors are compared to those from individual ones.

\noindent \\ \textbf{Clustering:} Additional analysis profiles players and teams into three clusters each, based on explanatory factors, using k-means clustering. Clusters are then manually labeled based on the average values of performance factors in each cluster. Win frequencies within these clusters are identified. Cross-tabulation shows how often teams with a majority of a certain player type win. Comparing these win rates across player and team clusters supports testing \textbf{H1}.

\section{Results}\label{sec:results}
Three main results are illustrated in this paper -- 
\subsection{Individual Acquisition but Collective Cooperation} \label{sec:result1}
Exploratory factor analysis on individual and team collective metrics reveals latent factors related to \emph{acquiring} versus \emph{sharing} in individuals, and \emph{cooperative} versus \emph{noncooperative} behaviors in teams. Scree analysis identifies the optimal number of factors as $N=2$ for both levels. Based on factor loading values, individual factors reflect either resource \emph{acquisition} or \emph{sharing}, while team factors represent either \emph{cooperation} or \emph{noncooperation}. Table~\ref{tab:factor_loadings} shows these loading values. Using these latent factors, classification models are trained to predict game outcomes. Feature importance scores reveal a dynamic: individuals tend to win through resource acquisition, while teams win through cooperative behaviors. At the individual level, acquisition is more critical [score = 0.68], however at the team level, cooperation and sharing [score = 0.82] are more influential than noncooperative strategies.

\begin{table*}[!htbp]
\centering
\caption{Factor loading values of latent factors for individual and collective team metrics are shown. For individual metrics, gray-shaded ones benefit others more than the player, while non-shaded ones reflect self-resource acquisition. Factor 1 shows high positive correlation with resource acquisition and low negative with sharing, and vice versa for Factor 2, distinguishing them as acquiring and sharing factors. For collective metrics, gray shading indicates resistance to the flow or centralization of resources and assistance. Factor 1 shows low positive correlation with centralization and high negative with resource acquisition (average gold and experience), while Factor 2 shows higher positive correlation with centralization metrics. Thus, Factor 1 and Factor 2 are identified as cooperative and noncooperative factors respectively.}
\label{tab:factor_loadings}
\begin{tabular}{cccccc}
\toprule
\textbf{Individual metrics} & \textbf{Factor 1} & \textbf{Factor 2} & \textbf{Collective metrics} & \textbf{Factor 1} & \textbf{Factor 2} \\
\midrule
Experience per minute & 0.7226 & -0.3297 & Avg. experience per minute & -0.8404 & -0.1311 \\
\cellcolor{gray!15} Vision score per minute & \cellcolor{gray!15}-0.2679 & \cellcolor{gray!15}0.9298 & \cellcolor{gray!15}Team in-degree centrality & \cellcolor{gray!15}0.1518 & \cellcolor{gray!15}0.4303 \\
Player in-degree & 0.7074 & -0.0432 & Avg. vision score per minute & -0.3414 & -0.1032 \\
\cellcolor{gray!15}Player out-degree & \cellcolor{gray!15}0.0305 & \cellcolor{gray!15}0.5013 & \cellcolor{gray!15}Team out-degree centrality & \cellcolor{gray!15}0.0696 & \cellcolor{gray!15}0.5859 \\
Player gold per minute & 0.9749 & -0.2159 & Avg. gold per minute & -0.9837 & -0.0699 \\
 & & & \cellcolor{gray!15}Effective graph resistance & \cellcolor{gray!15}0.5320 & \cellcolor{gray!15}0.5152 \\
\bottomrule
\end{tabular}
\end{table*}

\begin{figure}[!htbp]
    \centering
    \includegraphics[scale=0.7]{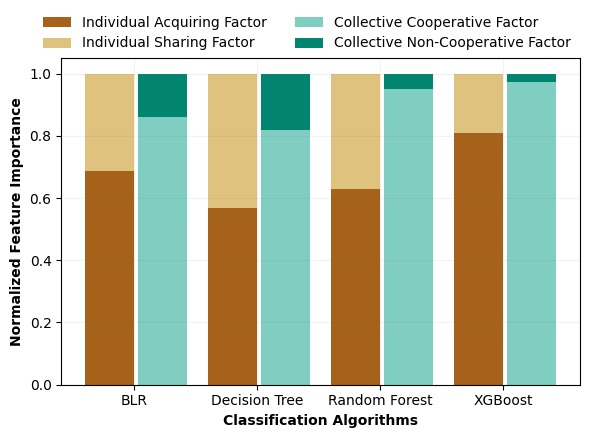}
    \caption{Feature importance values for individual and team collective performance factors across different classification algorithms. Across all classification models, at the individual level, the factor which represents acquisition of resources for a player is more crucial in winning the game rather than sharing. However, at the collective team level, cooperative behaviors such as resource sharing takes precedence over non-cooperative behaviors such as centralization of resources.}
    \label{fig:relative_importance_scores}
\end{figure}
 
 \subsection{Collective Intelligence Outperforms Individual Intelligence} \label{sec:result2}
As outlined by the previous result, individuals tend to focus on the acquisition of in-game resources but cooperative behavior remains paramount at the collective team-level. Evaluation scores of different classification algorithms to predict the winning outcome based on such individual and collective team-level data reveal that collective intelligence outperforms individual intelligence. Table~\ref{tab:evaluation_scores} presents the classification results of different machine learning algorithms based on the individual and collective performance factors. 

\begin{table}[htbp!]
\centering
\caption{Evaluation results of different classification algorithms to predict the winning outcome based on individual and collective performance factors.}
\label{tab:evaluation_scores}
\begin{tabular}{@{}llccc@{}}
\toprule
\multirow{2}{*}{\textbf{Analysis}} & \multirow{2}{*}{\textbf{Classification}} & \multicolumn{3}{c}{\textbf{Evaluation Metrics}} \\
\cmidrule(l){3-5}
\textbf{Level} & \textbf{Algorithms} & \textbf{Accuracy} & \textbf{F1-Score} & \textbf{AUC} \\
\midrule
\multirow{4}{*}{Individual} & BLR & 0.7036 & 0.70 &0.78 \\
& Decision Tree & 0.6203 & 0.62 & 0.62 \\
& Random Forest & 0.7115 & 0.71 & 0.78\\
& XGBoost & 0.7112 & 0.71 & 0.78 \\
\midrule
\multirow{4}{*}{Collective} & BLR & 0.8734 & 0.87 & 0.95 \\
& Decision Tree & 0.8260 & 0.83 & 0.82 \\
& Random Forest & 0.8734 & 0.87 & 0.95  \\
& XGBoost & 0.8730 & 0.87 & 0.95 \\
\bottomrule
\end{tabular}
\end{table}

Furthermore, k-means clustering is used to profile teams and individual players based on these factor scores. The elbow method identified the optimal number of clusters for both individuals and teams to be $N=3$. Such clusters for individual players and teams is further labeled based on the average distribution of performance factors in each cluster. That is, individual player clusters are labeled as \textit{acquiring, sharing} and \textit{average}, while team clusters are labeled as \textit{cooperative, non-cooperative} and \textit{average}. From the experimental dataset containing 31,055 matches, there is a total of 310,550 players (10 players in each match) and a total of 62,110 teams (2 teams in each match). Table~\ref{tab:winning_rate_for_labels} shows the distribution of winning rate across different clusters of teams and individual players. Additionally, we perform cross-tabulation analysis on the distribution of winning rate in different team clusters having a majority of different types of players in Table~\ref{tab:cross_analysis_winning_rate}. Table~\ref{tab:consolidated_table} presents a consolidated view of Table ~\ref{tab:winning_rate_for_labels} and Table~\ref{tab:cross_analysis_winning_rate} comparing the winning rate of cooperative teams having a majority of average or sharing players (i.e. medium to low-performing individuals) with non-cooperative teams having a majority of average or acquiring players (i.e. medium to high-performing individuals). 

\begin{table}[!htbp]
\centering
\caption{Distribution of winning rate across different clusters of teams and players. Cooperative teams are more likely to win games (about 89\%) compared to non-cooperative ones (20\%). Also, resource acquiring players are less likely to win the game than a cooperative team (i.e. 72\% vs 89\%). A resource sharing player has greater likelihood of winning the game than a non-cooperative team (i.e. 52\% vs 20\%).}
\label{tab:winning_rate_for_labels}
\begin{tabular}{ccccc}
\toprule
\textbf{Labels} & \textbf{Games} & \textbf{Wins} & \textbf{Losses} & \textbf{Win Rate}\\
\midrule
Cooperative Team & 28,517 & 25,535 & 2,982 & 0.8954 \\
Non-cooperative Team & 26,498 & 5,348 & 21,150 & 0.2018 \\
Average Team & 7,095 & 172 & 6,923 & 0.0242 \\
Overall & 62,110 & 31,055 & 31,055 & 0.5000 \\
\midrule
Acquiring Player & 112,862 & 81,672 & 31,190 & 0.7236 \\
Sharing Player & 53,413 & 28,077 & 25,336 & 0.5257 \\
Average Player & 144,275 & 45,526 & 98,749 & 0.3156 \\
Overall & 310,550 & 155,275 & 155,275 & 0.5000 \\
\bottomrule
\end{tabular}
\end{table}

\begin{table*}[htbp!]
\centering
\caption{Cross-tabulation analysis of winning rates in cooperative/non-cooperative teams with majority ($N\geq3$) of the acquiring, sharing or average players within the team. The rows indicates values for cooperative and non-cooperative teams whereas the column labels acquiring, sharing and average denote matches with majority of such of players in that team. Strikingly, cooperative team with a majority of resource acquiring players is the most likely to win the game.}
\label{tab:cross_analysis_winning_rate}
\resizebox{0.98\textwidth}{!}{
\begin{tabular}{@{}lccccccccc@{}}
\toprule
\multirow{2}{*}{\textbf{Teams}} & \multicolumn{3}{c}{\textbf{Matches}} & \multicolumn{3}{c}{\textbf{Wins}} & \multicolumn{3}{c}{\textbf{Winning Rate}} \\
\cmidrule(lr){2-4} \cmidrule(lr){5-7} \cmidrule(lr){8-10}
 & \textbf{Acquiring} & \textbf{Sharing} & \textbf{Average} & \textbf{Acquiring} & \textbf{Sharing} & \textbf{Average} & \textbf{Acquiring} & \textbf{Sharing} & \textbf{Average} \\
\midrule
Cooperative & 18,079 & 0 & 2,376 & 17,032 & 0 & 1,831 & 0.9421 & - & 0.7706 \\
Non-cooperative & 1,366 & 2 & 18,950 & 747 & 1 & 2,499 & 0.5469 & 0.500 & 0.1318 \\
\bottomrule
\end{tabular}}
\end{table*}

\begin{table}[!htbp]
\centering
\caption{Winning rates of cooperative teams having a majority of low-to-medium skilled individuals compared with non-cooperative teams having a majority of medium-to-high skilled individuals. Overall, low-performing individuals in a good team can out-compete high-performing individuals in a bad team (81\% vs 21\%)}
\label{tab:consolidated_table}
\begin{tabular}{cccc}
\toprule
\textbf{Team Labels} & \textbf{Games} & \textbf{Wins} & \textbf{Win Rate}\\
\midrule
Coop. (T) + Low-skilled (I) & 10,438 & 8,503 & 0.8146 \\
Non-coop. (T) + High-skilled (I) & 26,496 & 5,347 & 0.2018 \\
\bottomrule
\end{tabular}
\end{table}

\begin{figure}[!htbp]
    \centering
    \includegraphics[scale=0.9]{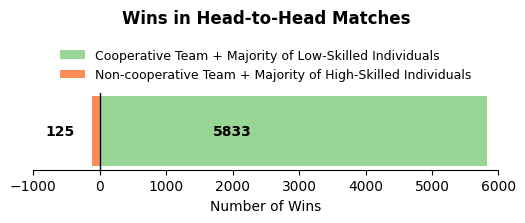}
    \caption{Frequency of wins for cooperative teams having a majority of average-to-low skilled individuals and non-cooperative teams having a majority of low-to-average skilled individuals when they face each other.}
    \label{fig:h2h_matchups}
\end{figure}

Moreover, we compare the winning ratio of match-ups when cooperative teams with a majority of low-to-average skilled individuals face-off against non-cooperative teams having a majority of average-to-high skilled individuals as shown in Figure~\ref{fig:h2h_matchups}. In such matches, cooperative teams win about 98\% of the games. These analyses further signify that collective intelligence outperforms individual intelligence in winning games.

\begin{figure}[!htp]
    \centering
    \includegraphics[width=\columnwidth]{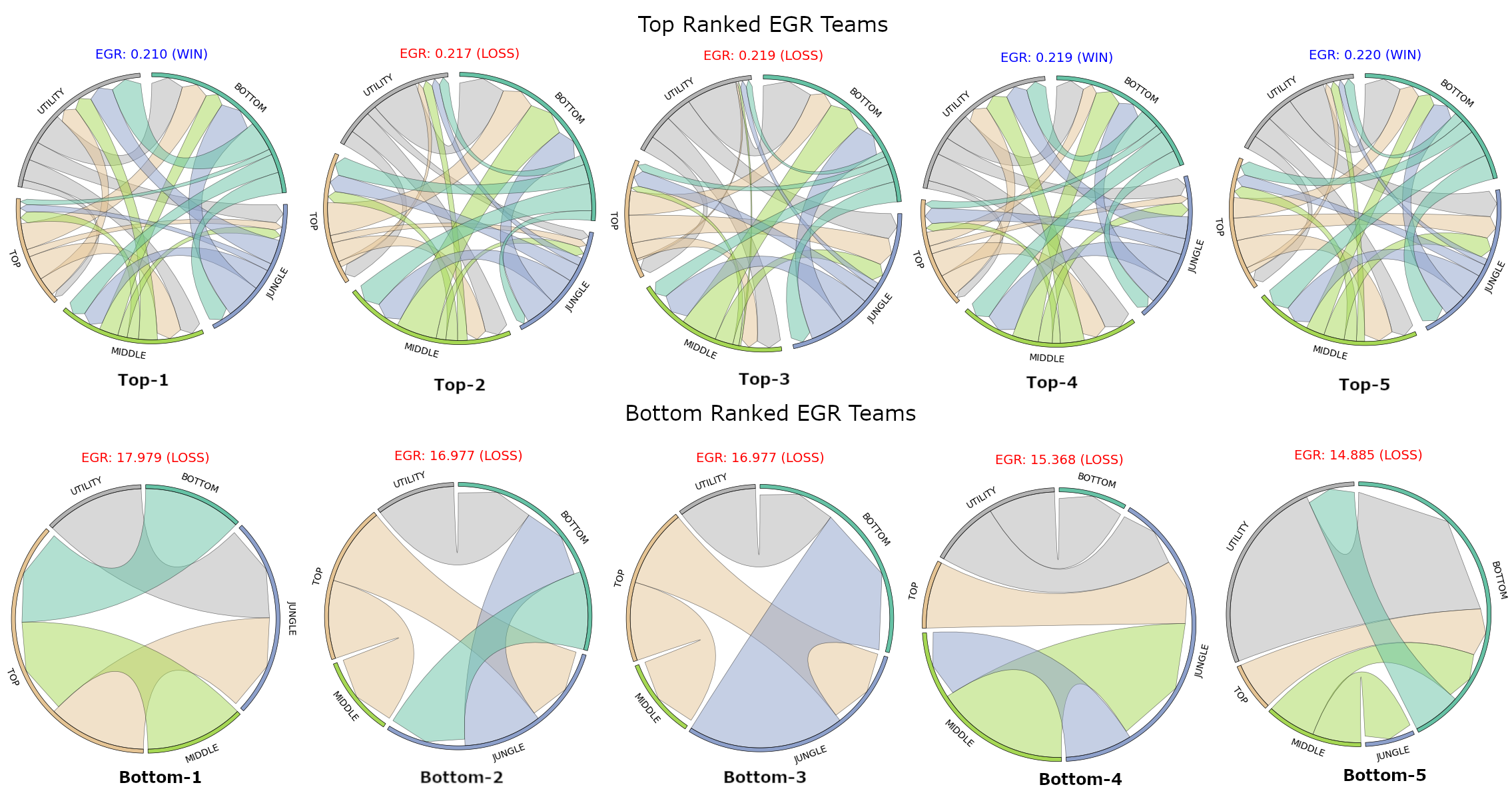}
    \caption{Chord diagrams based on data of graph networks are used here as a visual approach for the communication and analytics of collective intelligence within \emph{League of Legends}. The team interactions between the players are visualized for the ones with the top-5 and bottom-5 values of effective graph resistance (ranked from lowest to highest). For each team, the result of match being a loss or win is indicated. There are five nodes in each chord plot signifying unique player positions in the game. The thickness of the chords signify the weight of successful interactions (assistance) between players in a  team relative to the overall performance of the team. Teams with top-ranked EGR (lower) values exhibit multi-directional flow of interactions among players in a team while teams with bottom-ranked EGR (higher) values exhibit minimum interactions among each other.}
    \label{fig:visualizations}
\end{figure}

Last but not least, a visual approach is also used to demonstrate the superior performance of collective intelligence in Figure~\ref{fig:visualizations}. For this, we pick up one of the strongest indicators of collective intelligence in Table~\ref{tab:factor_loadings}, i.e. the effective graph resistance, to rank the teams and their interactions. The top-5 and bottom-5 teams are presented to visually inspect the nature of the interactions. As shown in Figure~\ref{fig:visualizations}, the top-5 (lowest EGR) teams show diverse interactions between the players and most of those win the match. In contrast, the bottom-5 (highest EGR) teams have more limited and isolated interactions between the players and all of those result in a loss. 

\subsection{Highly-skilled players exhibit higher levels of collective intelligence} \label{sec:result3}
\begin{figure}[!htbp]
    \centering
    \includegraphics[width=\columnwidth]{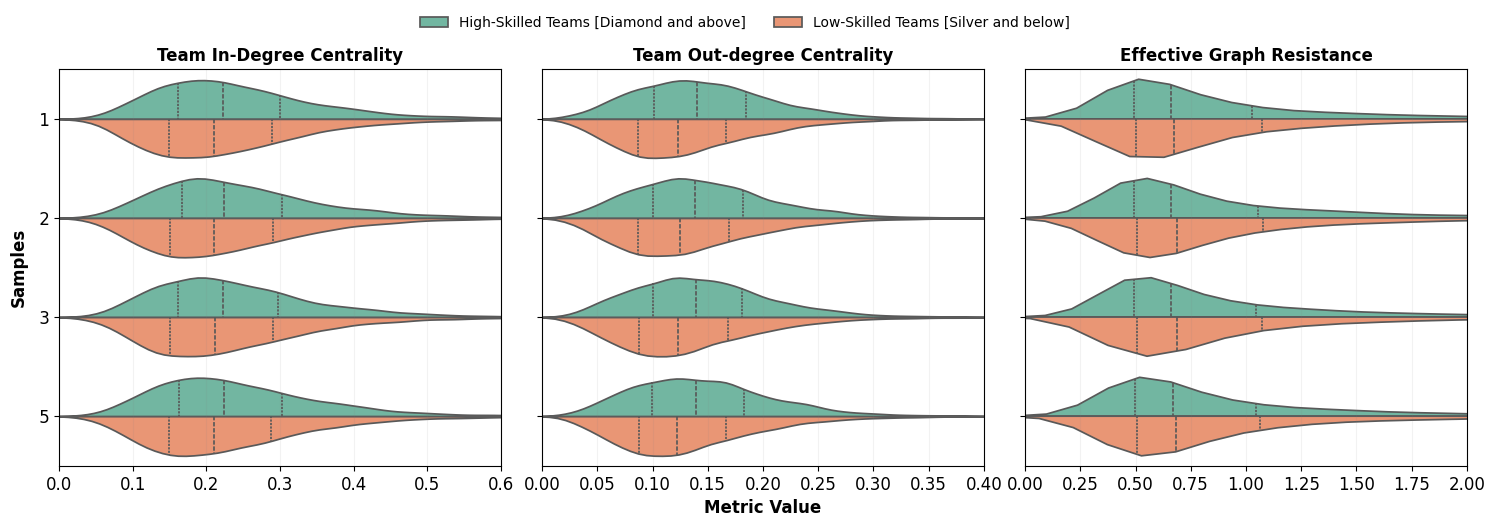}
    \caption{Teams in higher ranks (Diamond and above) demonstrate greater collective intelligence  for winning teams, as indicated by lower effective graph resistance, but higher in-degree and out-degree centrality, compared to lower-ranked teams (Silver and below). Randomly sampled winner and loser teams (5 times), sample size = 8000 }
    \label{fig:ranking_violin_plots}
\end{figure}

To test for \textbf{H2}, we investigated the association of graph-based collective intelligence metrics with the average rank of a team. Players who excel in the game are signified by a higher positioning in the ranking ladder. We find that teams containing highly-skilled players, based on their average team rank, exhibit moderately higher levels of collective intelligence (signified by lower values of effective graph resistance) compared to less-skilled teams. When teams are partitioned into two groups i.e highly-skilled and less-skilled, based on the values of average team rank sampled at size N=8000 teams with even distribution of wins and loss for 5 random samples, statistically significant differences emerge between the groups with respect to collective intelligence indicators. Specifically, the distributions of effective graph resistance, in-degree centrality, and out-degree centrality differ significantly across the two groups (two-tailed t-test, p < 0.05~\cite{supplementarydataset}, see Figure~\ref{fig:ranking_violin_plots}). However, less-skilled teams exhibit lower degrees of centrality compared to highly-skilled teams. Such can be the case as highly-skilled teams are more aware of player positions and roles, and thus tend to focus resources towards specific players that can win the game.

\section{Discussion}\label{discussion}
This paper unravels collective intelligence in \emph{League of Legends}. The first hypothesis, \textbf{H1:collective intelligence outperforms individual intelligence} has been analyzed and demonstrated using classification models, clustering algorithms and cross-tabulation analysis. Thus, teams prioritizing collective goals win more often than those relying on individual skill. Even low-performing players in cooperative teams outperform high-performing individuals in noncooperative ones. Although \emph{League of Legends} is designed as a collaborative team-based game, individuals are often matched up with random players which often shifts the emphasis from required coordinated team play to individual performance. The novel contribution of this work lies in the visual representation and analysis of team networks to re-iterate the importance of collective efforts over isolated individual skills. Visual analysis highlights complex interactions and low effective graph resistance as key indicators of collective intelligence. We are the first to use effective graph resistance for modeling team networks in \emph{League of Legends}, offering new insights through visual analytics and communication.  Building on Mora-Cantallops and Sicilia~\cite{mora2019team}, who found resource centralization harms performance, our study extends this by incorporating all assist types, including \emph{positional map-pressure}. Compared to the controlled settings of Kim et al.~\cite{kim2015work}, our large-scale dataset of anonymous players provides broader validity.

For the second hypothesis, \textbf{H2: highly-skilled players exhibit higher levels of collective intelligence compared to less-skilled players}, we presented statistically significant evidence showing that highly-skilled teams and players (signified by the ranking system) exhibit higher levels of collective intelligence compared to less-skilled players via two-tailed t-test.

In our results, we find that non-cooperative teams having a majority of sharing (or low-performing) individuals win 50\% of the games (see Table~\ref{tab:cross_analysis_winning_rate}). However, the observed number games for such non-cooperative teams is significantly low ($N=2$). Thus, the winning rate statistic for this particular case can be misleading. To remedy such issues, a larger dataset could be collected and analyzed. As we demonstrate a dominating winning ratio of cooperative teams with low-performing individuals over non-cooperative teams with high-performing individuals in head-to-head matches, it would be interesting to identify the distribution of such winning ratio across different \emph{tier} rankings. Additionally, a compelling study could be carried out whereby random, specific individuals are profiled for matches where they display exceptional performances in non-cooperative teams as well as mediocre performances in cooperative teams, and how often such displays can be attributed to wins, and their significance in climbing up the ranking ladder. As Krause et al.~\cite{krause2011swarm} and Santos et al.~\cite{santos2012role} advocate diversity for increased collective performance, the diversity in the types of \emph{champions} picked in game, can be studied to extend this work.

The phenomenon of collective intelligence in multiplayer online gaming environments merits investigation into diverse gaming genres, for instance in First-Person Shooter (FPS) games, Massive Multi-player Online Role Playing Games (MMORPGs). Unlike FPS games (which mostly focus on individual reflex and mechanics) and MMORPGs (in which games can span over a longer time period), the authors chose to study collective intelligence in MOBA games due to their well-defined team structures, player roles, resource and objective trade-offs that bear similarities to collective actions in real-world problems. Specifically, the authors chose to conduct this study in \emph{League of Legends} because -- (a) it is the most popular MOBA game played worldwide, and (b) features API endpoints that allow access to large-scale structured data. The framework that is presented in this work can also be used in other MOBA games like DOTA 2 to quantify the levels of outperformance. However, implementation will require tweaking the source code. 

Our results highlight the contrast between individual behavior when optimizing for self versus for the collective. This dynamic also appears in sustainable infrastructure, where individuals may choose cheaper non-sustainable products over costly eco-friendly ones that benefit society long-term~\cite{asikis2021value}. The study suggests individuals may be guided to prioritize collective altruism in complex social systems. Such collective behavior is vital not only in online games but also in real-world contexts. For instance, it can help combat misinformation in online networks and support fairer outcomes in socio-economic systems like consumer markets and governments by preventing monopoly and power abuse.

\section{Conclusion}\label{sec:conclusion}
In this paper, we explored the presence of collective intelligence among players in the MOBA game, \emph{League of Legends} using quantitative methods and a visual approach to knowledge discovery. Our findings conclude that collective intelligence characterized by sharing of resources, flow of information and cooperative behavior outperforms individual intelligence. Our results reiterate the paradoxical imperative for collaboration and cooperation in highly competitive environments for achieving optimal outcomes. Further work complementing this research can be done by testing for the performance of collective intelligence and individual intelligence in real-world applications beyond gaming environments.

\section*{Acknowledgements}{This work is funded by a UKRI Future Leaders Fellowship (MR\-/W009560\-/1): \emph{Digitally Assisted Collective Governance of Smart City Commons--ARTIO}'.}

\section*{Code, Dataset and Supplementary Materials} \label{sec:code}
Relevant source code used in the analyses outlined in the paper is available at \noindent{\url{https://github.com/TDI-Lab/Collective-Intelligence-LoL}. Supplementary materials on game details and collected dataset is available under the corresponding Zenodo record~\cite{supplementarydataset}.

\bibliographystyle{unsrt}
\bibliography{sample}

\end{document}